\documentclass[12pt,a4paper]{article}
\usepackage{amssymb,axodraw}
\usepackage{amssymb}
\newcommand{\beq}{\\[0.5ex]\begin{equation}}
\newcommand{\eeq}{\\[0.5ex]\end{equation}}

\begin{document}
\title{Memory Effects in Long-Time Lindblad Motion}
\author{Klaus Dietz\thanks{Permanent Address:Physics Department,University of Bonn, 53115 Bonn, Germany}\\
Max Planck Institut f\"ur Physik komplexer Systeme \\
01187 Dresden, Germany}
\maketitle
\begin{abstract}
   It is shown that the Lindblad equation accounts for memory
   effects. That is to say, Lindblad operators can be constructed
   in a natural manner such that a memory term appears in the
   asymtotic $\left(time \to \infty \right)$ region; at the same time
   the expectation values depend on the initial state. Furthermore a
   procedure to extend the Lindblad equation to an equation of motion
   for an ideal Bose 'gas' of 'particles',i.e. systems with non-trivial
   internal structure,is described.Initially in some quantum state
   this collection of 'particles' will asymptotically turn into an
   equilibrium ensemble whose probability distribution is determined
   by the Lindblad operators building the dissipative part of the
   equation of motion. 
\end{abstract}

\noindent
A)The Lindblad generalisation \cite{lindblad} of Schroedinger motion to
   dissipative motion hinges, apart from technical assumptions, only
   on very general physical notions:\\
\begin{itemize}
\item[i)]The abelian group of unitary Schroedinger motion generated by the
Hamiltonian generalizes to a set of abelian semigroups characterized
by a collection of operators ${V_J}$,the Lindblad operators. The
semigroup structure - the non-existence of time-reversed motion -
accounts for absorption.\\
\item[ii)]Complete positivity, interpreted physically, means that positive
motions in Hilbert space of states (system 1) can be extended to a positive
motion in the product space resulting by adjoining a second Hilbert
space (system2), a construction leading to entanglement of both
systems.
\end{itemize}
It should be noted that in the derivation of the Lindblad equation
concepts used in the derivation of master equations - for instance the
decomposition of the space of states in a product  $'system' \otimes
'bath'$ - do not play any role, at no place. Nor is there any
conceptual relation to open systems.The relation of master-equations
for open systems and Lindblad equations has been clarified to some
extent \cite{haake} \cite{davies} \cite{alicki}
 \cite{dumcke}\cite{strunz} \cite{gorini} and
shown to be controlled by relative scales.\\
We take up this observation and consider a system whose degrees of
freedom interact with scale-dependent Hamiltonians and look for
stationary states evolving from given initial states,i.e. we construct
maps ($\rho$ is the density operator of our system)
\beq
\tau\left(V\right):\quad \varrho |_{t=0}\quad\mapsto\quad\varrho |_{t=\infty}
\eeq
and discuss their dependence on the Lindblad operators $V_J$ which
together with the Hamiltonian are supposed to differ for different
scales - time scales,energy scales etc.\\
We present an explicit construction of asymptotic stationary states
which will be seen to contain memory terms.\\

\noindent
B)In this section we consider the case of only one Lindblad operator
$V$ and write the Lindblad equation of motion
\beq
\dot{B} = i ~ [H,B] + V^{+}BV -\frac{1}{2}[V^{+} V, B]_{+}
\eeq
where $H$ is the Hamiltonian and $B$ an observable.$\footnote{V is
  assumed to be invertible; B,H
  are bounded operators acting in a
  separable Hilbert space for which the Lindblad equation has
  been proved.This fact allows us to use interchangeably the notions
  'operator' and 'matrix' and treat the question of dimensions -
  finite or infinite -  in a rather cavalier way.}$\\
Using the polar decomposition (U is a unitary operator)
\beq
V=U\sqrt{V^{+}V}
\eeq 
\newpage
\noindent
we rewrite (2) as ( note that the assumption of unitarity of $U$
excludes zero-modes in $V^{+}V$ )
\begin{eqnarray}
\lefteqn{\frac{1}{\sqrt{V^{+}V}}\dot{B}\frac{1}{\sqrt{V^{+}V}} =} \nonumber\\[2ex] 
&& i \frac{1}{\sqrt{V^{+} V}}~[H,B]~\frac{1}{\sqrt{V^{+} V}}+ U^{+}BU \nonumber\\[2ex]
&& - \frac{1}{2}\left(\frac{1}{\sqrt{V^{+}V}}B\sqrt{V^{+}V} + 
\sqrt{V^{+}V}B\frac{1}{\sqrt{V^{+}V}}\right) \nonumber\\[2ex]
\end{eqnarray}
The observation (see below) that
\beq
W:=\frac{1}{V^{+}V}
\eeq
is a (non-normalized) probability distribution leads us to the
physically plausible assumption
\beq
\label{3}
W=W(H,\ldots)
\eeq
where the dots indicate further observables commuting with
H. Tracing the equation of motion we immediately see that the trace of
the rhs of the equation of motion vanishes identically and hence
\beq
\label{1}
tr(\dot{B}W) = 0
\eeq
or
\beq
tr(BW)=const
\eeq
( $\dot{W}=0$ since $W$ depends only on conserved
quantities). Needless to say we tacitly assume W to be traceclass.\\

\noindent
In \cite{dietz} we have demonstrated the following asymptotic form for
B
\begin{itemize}
\item[i)]Irreducible V
\beq
B |_{t=\infty} = b(\infty)\mathbb{I}
\eeq
where $\mathbb{I}$ is the unit operator in $\mathfrak{H}$.
\item[ii)] Reducible V, i.e.
\beq
V = \sum_{\alpha} \oplus V_{\alpha}
\eeq
where the $V_{\alpha}$ are matrices in orthogonal subspaces
$\mathfrak{H}^{\alpha}$ of $\mathfrak{H}$, yield
\beq
B |_{t=\infty} = \sum_{\alpha} \oplus
b^{\alpha}(\infty)\mathbb{I}^{\alpha}.
\eeq
\end{itemize}
In the following we consider only the irreducible case and derive 
\beq
B |_{t=\infty} = \frac{tr(B |_{t=0}W)}{tr(W)}\mathbb{I}
\eeq
The expectation value of the asymptotic configuration then is 
\beq
\label{5}
< B |_{t=\infty} > = tr(B |_{t=0}\varrho_{0}) = 
\frac{tr(B |_{t=0}W)}{tr(W)}
\eeq
for all states $\varrho_{0}$, i.e. the expectation value is
independent of the initial state, no memory effects are present.We see that
\beq
P_{W} = \frac{W}{tr(W)}
\eeq
is a normalized probability distribution derived from the Lindblad
operator $V$.Translating this result into the Schroedinger picture we derive that
any initial state $\varrho_{0}$ tends to $P_{W}$ for $t \to \infty$, i.e.
\beq
\tau\left(V\right):\quad\varrho_{0}\quad \mapsto\quad \varrho_{0} |_{t = \infty} =P_{W}
\eeq
for all initial states $\varrho_{0}$.\\

We now turn to the question of memory effects. To show that they can
be incorporated we extend the Lindblad equation, without changing
its formal content, to an equation of motion for quantum subsystems
separated, e.g. by scales, from the system built up by these
subsystems. As an example we could take a molecule: the subsystems are
spanned by the states corresponding to the inner degrees of freedom of
the atoms composing the molecule, the system - the molecule - is built
up by the atomic states of outer shells.\\
To realize this construct we endow the input matrices V and H with a
direct product structure and, to simplify matters, choose the
ansaetze
\beq
V = \left(\; \tilde{V}_{i k}\mathbb{I}\; \right)\sqrt{n}
\eeq
\beq
H = \left(\; \tilde{H}_{i, k}\mathbb{H}\; \right)
\eeq
\beq
\tilde{V}_{i k},\tilde{H}_{i k} \in \mathbb{C}
\eeq
where $\mathbb{I}$ is the $n \times n$ unit matrix and $\mathbb{H}$ is a
$n \times n$ matrix sub-Hamiltonian, identical for all sites $\left(
i,k \right)$, i.e.$\tilde{V}$ and $\tilde{H}$ are matrices with $n \times n$ matrix
valued entries indexed by $\left(i,k \right)$. Our ansatz for V
guarantees that,in the terms of our example, the Lindblad operator
leaves the inner degrees of freedom unaffected. The observable B is written as a
matrix of $n \times n$ matrices $ B_{i k}$ 
\beq
B = \left(\; B_{i k}\; \right)
\eeq
The probability distribution is then
\beq
W =\left( \left(\tilde{V}^{+} \tilde{V} \right)^{-1} \otimes \mathbb{I}\right) =:\left( \tilde{W}_{i
  k} \otimes \mathbb{I} \right),
\eeq
V has the polar decomposition 
\beq
V = U~\sqrt{V^{+} V} =
\left(\tilde{U} \otimes \mathbb{I}\right)\sqrt{\tilde{V}^{+}\tilde{V}
  \otimes \mathbb{I}}. 
\eeq
Tracing the equation of motion with respect to the indices $\left(i,k
\right)$ then yields instead of (\ref{1}) 
\beq
\tilde{Tr}\left( \dot{B} W\right) =
\mathbb{H} \tilde{Tr}\left(\tilde{W}\tilde{H}B\right) -
\tilde{Tr}\left(\tilde{W}B\tilde{H}\right)\mathbb{H} 
\eeq
which leads to
\beq
\label{2}
\tilde{Tr}\left(B W \right) = C
+\int_0^{t}\left(\mathbb{H}\tilde{Tr}\left(\tilde{W}\tilde{H}B(t)\right)-
\tilde{Tr}\left(\tilde{W}B(t)\tilde{H}\right)\mathbb{H}\right)\,dt
\eeq
It has to be stated that taking the trace of this $n \times n$ matrix
we obtain a vanishing result
\beq
Tr\left(\dot{B} W\right) = tr_{n}\tilde{Tr}\left(\dot{B}W\right) = 0
\eeq
in accordance with equation (\ref{1}) .This is because we have assumed
\beq
0 = \left[W,H\right] = \left[\tilde{W},\tilde{H}\right] \otimes
\mathbb{H}
\eeq
and, thus
\beq
\left[\tilde{W},\tilde{H} \right] = 0.
\eeq
Following the derivation given in \cite{dietz} we find for the
asymtotic configuration
\beq
B |_{t=\infty} = b(\infty)~\left(\;\delta_{i k}\; \right)
\eeq
where $b(\infty)$ is now a $n \times n$ matrix which is read off 
\begin{eqnarray}
\lefteqn{b\left(\infty \right) =} \nonumber\\ 
&& \frac{1}{\left( \tilde{Tr}\tilde{W} \right)} \times \nonumber\\
&& \left(\tilde{Tr}\left(B|_{t=0}~W\right) +
\int_0^{\infty}\left(\mathbb{H}~\tilde{Tr}\left(\tilde{W}\tilde{H}B(t)\right)
- \tilde{Tr}\left(\tilde{W}B(t)\tilde{H}\right)\mathbb{H}\right)\,dt
\right) . \nonumber\\
\end{eqnarray}
We note the explicit appearance of a memory term.Calculating the
expectation value of $B|_{t=\infty}$ in some state $\varrho_{0}$
written as
\beq
\varrho_{(0)} = \left( \varrho^{(0)}_{i k} \right)
\eeq
where the $\varrho^{(0)}_{i k}$ are $n \times n$ matrices, we find
\beq
< B|_{t=\infty} > = \sum_{i}tr_{n}\left(\varrho^{(0)}_{i i}b(\infty)
\right)
\eeq
and observe that now the asymptotic expectation value does depend on
the initial state in concordance with the appearance of a memory
term.\\

So we have seen that a simple and intuitively clear generalisation of
the Lindblad equation to an equation for dynamical degrees of freedom
of subsystems leads to memory effects; the asymptotic subsystem
variables given in equations (\ref{2}) should be interpreted as the new
dynamical subsystem variables obtained from an asymptotic
averaging procedure over those degrees of freedom of the total system
living on 'lower' scales:the equations (\ref{2}) are clearly seen as
an elimination procedure for 'environment' variables separated into a
statistical average and a memory term.\\

\noindent
C)We now turn to the case of more than one, say N, Lindblad
operators. We take $N$ finite with the provision of eventually letting
$N \to \infty$ as certain physical models might require.The equation of
motion then reads
\beq
\dot{B} = i[H,B] + \sum_{J}\left(V_J^{+}BV_J -
\frac{1}{2}\left[V_J^{+}V_J,B\right]_{+}\right)
\eeq
We rewrite this equation as an equation operating in a direct sum of
identical spaces
\beq
\mathfrak{H}_{N} = \sum_{1 \to  N} \oplus \mathfrak{H}
\eeq
and define
\beq
B_N := B\mathbb{I}_N
\eeq
\beq
H_N := H\mathbb{I}_N
\eeq
\beq
V_N := \sqrt{N}\left( \; V_J\delta_{J K} \; \right)
\eeq
to arrive at
\beq
\dot{B}_N = i[H_N,B_N] + V_N^{+}B_NV_N - \frac{1}{2}\left[
  V_N^{+}V_N,B_N\right]_{+}.
\eeq
The polar decompositions
\beq
V_J = U_J\sqrt{V_J^{+}V_J}
\eeq
lead to the polar decomposition
\beq
V_N = U_N\sqrt{V_N^{+}V_N}
\eeq
where
\beq
U_N = \left( \; U_J\delta_{J K} \;\right)
\eeq
\newpage
\noindent
is unitary.
Employing the same procedure as above we have
\begin{eqnarray}
\label{4}
\lefteqn{\frac{1}{\sqrt{V_N^{+}V_N}}\dot{B}_N~\frac{1}{\sqrt{V_N^{+}V_N}}
  =} \nonumber\\[2ex]
&& i\frac{1}{\sqrt{V_N^{+}V_N}}[H_N,B_N]\frac{1}{\sqrt{V_N^{+}V_N}} + U_N^{+}BU_N \nonumber\\[2ex]
&& - \frac{1}{2} \left( \sqrt{V_N^{+}V_N}B_N\frac{1}{\sqrt{V_N^{+}V_N}}
+ \frac{1}{\sqrt{V_N^{+}V_N}}B_N\sqrt{V_N^{+}V_N} \right) \nonumber\\.
\end{eqnarray}
Assuming either independence of $W_{J}$ on J ( $U_{J}$ does depend on $J$ in
general) or, alternatively, $V_{J}$ positive and in analogy to (\ref{3})
\beq
W_J := \frac{1}{V_J^{+}V_J} = W_J\left( H,\ldots\right)
\eeq
and taking the total trace ($tr_N$ pertains to the matrix indices of
the $N \times N$ matrices introduced above, $tr_{\mathfrak{H}}$ to the
operators on $\mathfrak{H}$) we find
\beq
tr_{\mathfrak{H}} tr_N \left(\dot{B}_NW_N \right) = 0
\eeq
and thus
\beq
tr_{\mathfrak{H}}\left(\dot{B}\sum_{1\to N}W_J \right) = 0
\eeq
and
\beq
\label{6}
<B |_{t=\infty} > =
\frac{tr_{\mathfrak{H}}\left(B|_{t=0}W\right)}{tr_{\mathfrak{H}}\left(W\right)}
\eeq
\beq
W = \sum_{J}W_J
\eeq
To illustrate this result we take $V_{J}$ positive and assume the following specific ansatz
\beq
W_J = \left( V^{+}V \right)^J
\eeq
\newpage
\noindent
$J = 1,2,\ldots$ and put
\beq
V = \exp{\left(-\frac{\beta H}{2}\right)}
\eeq
The expectation value for B reaches asymptotically
\beq
\label{7}
< B|_{t=\infty} > =
\frac{1}{\bar{N}}tr_{\mathfrak{H}}\left(\frac{B|_{t=0}}{\exp{\left(\beta H\right)}-1}\right)
\eeq
which is simply the expectation value of the 'particle' observable B
in an ideal Bose 'gas' of, in the average, $\bar{N}$ 'particles' at
inverse temperature $\beta$; 'particle' is just a more intuitive name
for the physical object dubbed 'system' up to now.\\
\noindent
This interpretation deserves further clarifications. To this end we
reformulate equation (\ref{4}) as an equation in Fock space $\mathfrak{H}_F$,
aiming at the Bose nature of the ideal gas to be introduced.We
define
\beq
\mathfrak{H}_F := \sum_{J}\oplus \mathfrak{H}^{\otimes J}
\eeq
and
\beq
V_F := \sum_{J}\oplus V^{\otimes J}
\eeq
so that
\beq
W_F = V_F^{+}V_F = \sum_{J}\oplus \left(V^{+}V\right)^{\otimes J}.
\eeq
In the product space $\mathfrak{H}^{\otimes J}$ we select as physically
relevant states symmetric states which we take as superpositions of
symmetric 'system' product states - we introduce many-'particle' boson
states .The observable $B$ is extended to a symmetrically operating
operator \\
\beq
B_F = \sum_J \oplus \left( \ldots \otimes \mathbb{I} \otimes B \otimes \ldots
\right)
\eeq
where B stands consecutively on all positions of the J-fold product.\\
The expectation value at $t=\infty$ is, in strict analogy to (\ref{5}) and
(\ref{6})\\
\beq
< B_F|_{t=\infty}>=\frac{tr_{\mathfrak{H}_F}\left(B_F|_{t=0}W_F
  \right)}{tr_{\mathfrak{H}_F}\left(W_F\right)}
\eeq
where the trace is now to be calculated with a symmetric product basis
in $\mathfrak{H}^{\otimes J}$ for all J. Computing this trace one
encounters disconnected terms (matrix elements now pertain to
$\mathfrak{H}$) 
\beq
\sum_{i_1,\ldots,i_L} <i_1|BV^{J_1}|i_1><i_2|V^{J_2}|i_2> \ldots
<i_L|V^{J_k} |i_L> 
\eeq
\beq
\sum_{l}J_l = J
\eeq
All these terms sum up to the same common factor in the nominator and
denominator - the connected cluster theorem - so that we reproduce (\ref{7})
with W$ _J = (V^{+}~V)^J$.
We conclude that any symmetric quantum many-particle state composed of
whatever complex quantum systems - a Bose many-particle state - is
transported by Lindblad motion into an equilibrium ensemble with a
probability distribution
\beq
W_{\mbox{$\scriptstyle Bose$}} = \frac{1}{W_{\mathfrak{H}} - 1}
\eeq
where
\beq
W_{\mathfrak{H}} = \frac{1}{V^{+}V}
\eeq
is an operator acting in the space of the system's states. This
derivation is a first step towards a dissipative quantum field
theory: the case of free fields, although we never explicitly introduced
this concept.I shall return to the extension to more complicated cases
in a forthcoming publication.\\

Acknowledgment:\\
It is a pleasure to thank Jan-Michael Rost, Max Planck Institut fuer
die Physik Komplexer Systeme, Dresden and Axel Schenzle,Sektion Physik,
Universitaet Muenchen for the stimulating ambiente during my stay at
these instutions where this work has been done.

\end{document}